\documentclass{article}

\usepackage{PRIMEarxiv}

\usepackage[utf8]{inputenc} 
\usepackage[T1]{fontenc}    
\usepackage{hyperref}       
\usepackage{url}            
\usepackage{booktabs}       
\usepackage{amsfonts}       
\usepackage{nicefrac}       
\usepackage{microtype}      
\usepackage{lipsum}
\usepackage{fancyhdr}       
\usepackage{graphicx}       
\graphicspath{{media/}}     
\usepackage{multirow}
\title{Rethinking Sales Lead Scoring with LLM-based Hierarchical Preference Ranking
\thanks{Email address: \href{cyzhang57@outlook.com}{cyzhang57@outlook.com}}}
\author{
  Chenyu Zhang, Yiwen Liu, Yin Sun, Xinyuan Zhang, Yuji Cao, Junming Jiao, Juyi Qiao \\
  Intelligent Business Team, Li Auto Inc. \\
  }
  
\begin{document}
\maketitle
\begin{abstract}
Sales lead conversion in  high-stakes domains (e.g., automotive, real estate) differs fundamentally from e-commerce recommendation due to prolonged decision cycles and multi-stage funnels.
Traditional lead scoring methods—rule-based scorecards, machine learning, or pointwise CTR models—face severe challenges: sparse supervision, a semantic gap in unstructured CRM logs, and inability to capture relative lead priority. While Large Language Models(LLMs) offer superior semantic understanding of customer interactions, general-purpose LLMs are ill-suited for lead ranking: they generate text rather than comparable scores, and lack alignment with the hierarchical priorities of sales funnels. 
We introduce an LLM-based discriminative framework for sales lead scoring, which supports joint modeling of structured CRM features and unstructured customer interactions.  On top of this framework, we propose HPRO (Hierarchical Preference Ranking Optimization), which augments sales lead scoring with a hierarchical preference ranking objective.    
HPRO employs a margin-aware Bradley-Terry formulation to transform sparse binary labels into dense, funnel-aware preference pairs, enabling lead scoring to leverage both pointwise and pairwise supervision.
Experiments on large-scale data from a leading NEV brand demonstrate state-of-the-art classification (AUC 0.8161) and ranking performance (+39.7\% precision among top-ranked leads). A 132-day online A/B test validates 9.5\% sales volume uplift, confirming real-world commercial impact. 
\end{abstract}

\keywords{Sales Lead Scoring \and  Large Language Models \and  Preference Optimization \and Learning to Rank}

\section{Introduction}
Lead scoring—ranking sales leads by conversion likelihood—is central to modern Customer Relationship Management (CRM)~\cite{winer2001framework,payne2005strategic}. The commercial importance of this task is reflected in substantial industry investment: platforms such as Salesforce Einstein, SAP CRM, and Act-On deploy machine learning-based scoring systems to prioritize high-potential leads~\cite{wu2024state,sap2022lead,acton2023ai}. Formally, lead scoring can be modeled as a prediction task akin to Click-Through Rate (CTR) estimation: given a lead's profile and interaction history, the objective is to estimate the probability of conversion~\cite{guenzi2020mastering,monat2011industrial}.

In standard recommendation scenarios, user feedback (clicks, views) arrives within seconds of an impression, providing dense supervision for model training~\cite{cheng2016wide,guo2017deepfm}. However, long-chain sales—prevalent in domains such as automotive, real estate, and enterprise B2B—involve extended decision cycles spanning weeks or months~\cite{verhoef2021digital,syam2018waiting}. Directly transplanting recommendation paradigms to this setting reveals fundamental incompatibilities.

A typical automotive purchase progresses through multiple funnel stages: Lead $\to$ Telemarketing $\to$ Store Visit $\to$ Test Drive $\to$ Order~\cite{wu2024state,conde2025necessary,boogar2019three}. This creates a critical challenge: the supervision signal is sparse not because positive labels are rare, but because the signal indicating conversion intent is diffused across a long decision horizon. A customer who completed a test drive and one who merely browsed the website are both labeled as ``non-converted'' if neither ultimately purchases—yet their intent levels differ substantially. Standard binary supervision conflates these cases, treating qualitatively different negatives as identical and discarding valuable process information~\cite{cadavid2018trends,mcdonnell2019evolution}.

Existing methods are ill-suited to this setting. Rule-based scorecards depend on expert heuristics that scale poorly~\cite{gonzalez2025relevance,mazur2025assessing}. Modern CRM platforms (e.g., Salesforce Einstein) have adopted machine learning methods that excel at structured features but optimize only terminal lock-in  labels~\cite{wu2024smart,couronne2018random}. Given the structural similarity to conversion prediction, deep CTR models—DeepFM~\cite{guo2017deepfm}, DCN~\cite{wang2017deep}, xDeepFM~\cite{lian2018xdeepfm}, AutoInt~\cite{song2019autoint}—offer a natural alternative. However, these models assume dense, immediate feedback and encode text as shallow embeddings, failing to capture semantic signals in sales dialogues.


Large Language Models (LLMs) offer superior semantic understanding and can interpret the rich, unstructured signals embedded in CRM logs—such as customer sentiment and implicit intent~\cite{brynjolfsson2025generative,noy2023experimental}. However, general-purpose LLMs are misaligned with lead ranking requirements. Recent benchmarks such as CRMArena~\cite{huang2025crmarena,huang2025crmarenapro} reveal that LLMs struggle with CRM tasks requiring structured reasoning over business objects and adherence to domain-specific rules. More fundamentally, generative models produce text rather than actionable scores—they lack mechanisms for comparable ranking outputs. This motivates a discriminative design: augmenting an LLM backbone with task-specific heads that output scalar scores amenable to ranking and calibration.

We propose HPRO (Hierarchical Preference Ranking Optimization), a framework that transforms sparse terminal supervision into dense, funnel-aware preference signals. The key insight is that while binary labels are uninformative about intermediate intent, the sales funnel itself encodes a natural preference hierarchy: test-drive customers are more engaged than phone-only inquiries. HPRO reformulates lead scoring as a Learning-to-Rank problem, constructing hierarchical preference pairs across funnel stages (e.g., Test Drive $\succ$ Store Visit $\succ$ Phone Call) and optimizing a margin-aware Bradley-Terry objective~\cite{bradley1952rank} where margins encode business priors about preference strength. This formulation draws inspiration from recent advances in preference optimization for LLMs~\cite{rafailov2023direct,hong2024orpo}, adapting these techniques to the sales domain.

This paper makes three contributions:
\begin{itemize}
    \item We introduce a discriminative LLM architecture that combines a semantic backbone with pointwise (calibration) and pairwise (ranking) heads, enabling joint modeling of tabular and textual features for sales lead scoring.
    \item We propose HPRO, which converts sparse binary supervision into hierarchical preference pairs derived from funnel stages. The margin-aware formulation explicitly injects domain knowledge about stage importance into the optimization objective.
    \item Experiments on  leads from a leading NEV brand show HPRO achieves state-of-the-art AUC of 0.8161 (+2.3\% over baselines). A rigorous 5-month online A/B test demonstrates \textbf{9.5\% sales volume uplift}, validating real-world commercial impact.
\end{itemize}

\section{Related Work}
\label{sec:headings}

\paragraph{Lead Scoring.}
Traditional lead scoring relies on rule-based scorecards~\cite{mcdonnell2019evolution} with manually assigned weights, prone to subjective bias~\cite{wu2024state}. Machine learning methods—Logistic Regression, Decision Trees~\cite{couronne2018random}, and ensemble models like XGBoost~\cite{mazur2025assessing,gonzalez2025relevance}—improved accuracy on structured CRM data. Modern platforms such as Salesforce Einstein deploy ML-based scoring~\cite{wu2024smart}, while some work explores funnel-aware analysis~\cite{conde2025necessary}. However, these approaches optimize only terminal labels, failing to capture semantic signals in dialogues or leverage intermediate supervision in long-chain sales.

\paragraph{Deep CTR Models.}
Deep CTR models—DeepFM~\cite{guo2017deepfm}, DCN~\cite{wang2017deep}, xDeepFM~\cite{lian2018xdeepfm}, AutoInt~\cite{song2019autoint}—have achieved strong results in recommendation by modeling high-order feature interactions. Graph-based extensions~\cite{wu2020comprehensive,yu2022graphfm} further capture entity relations. However, these methods assume dense, immediate feedback (clicks/views) unavailable in long-chain sales, and encode text as shallow embeddings, losing semantic nuance critical for distinguishing customer intent.

\paragraph{LLMs for CRM.}
LLMs have transformed CRM by enabling semantic understanding of unstructured interactions~\cite{brynjolfsson2025generative,noy2023experimental}. Yet benchmarks like CRMArena~\cite{huang2025crmarena} reveal that general-purpose LLMs struggle with structured reasoning over business objects and domain-specific rules. More fundamentally, generative LLMs lack mechanisms for producing ranking scores aligned with sales funnel hierarchy. This motivates our discriminative design: asLLR augments an LLM backbone with task-specific heads, while HPRO injects funnel-aware preferences via margin-based optimization.

\paragraph{Comparison with Preference Optimization Methods.}
Recent preference optimization methods such as DPO~\cite{rafailov2023direct} and ORPO~\cite{hong2024orpo} are designed to align generative models with human preferences in text generation. HPRO differs from these approaches in three aspects: it targets business ranking effectiveness rather than generation quality, constructs preference pairs automatically from funnel stages instead of manual annotations, and introduces hierarchical margins to encode domain-specific business priorities. HPRO also differs from traditional learning-to-rank methods, which typically assume either explicit relevance labels or uniformly weighted pairwise preferences, whereas our setting provides only sparse terminal labels and funnel-derived preference strengths.

\section{Methodology}

\subsection{Problem Formulation}
We formulate lead scoring as a \textit{Learning-to-Rank} problem augmented with probability calibration. Let $\mathcal{D} = \{(x_i, y_i)\}_{i=1}^N$ denote the dataset, where each sample $x_i = (T_i, L_i)$ comprises tabular features $T_i$ (e.g., demographics, behavioral statistics) and dialogue features $L_i$ (e.g., sales conversation transcripts). The binary label $y_i \in \{0, 1\}$ indicates whether final conversion (order lock-in) occurred.

\begin{figure*}[t]
    \centering
    \includegraphics[width=0.85\linewidth]{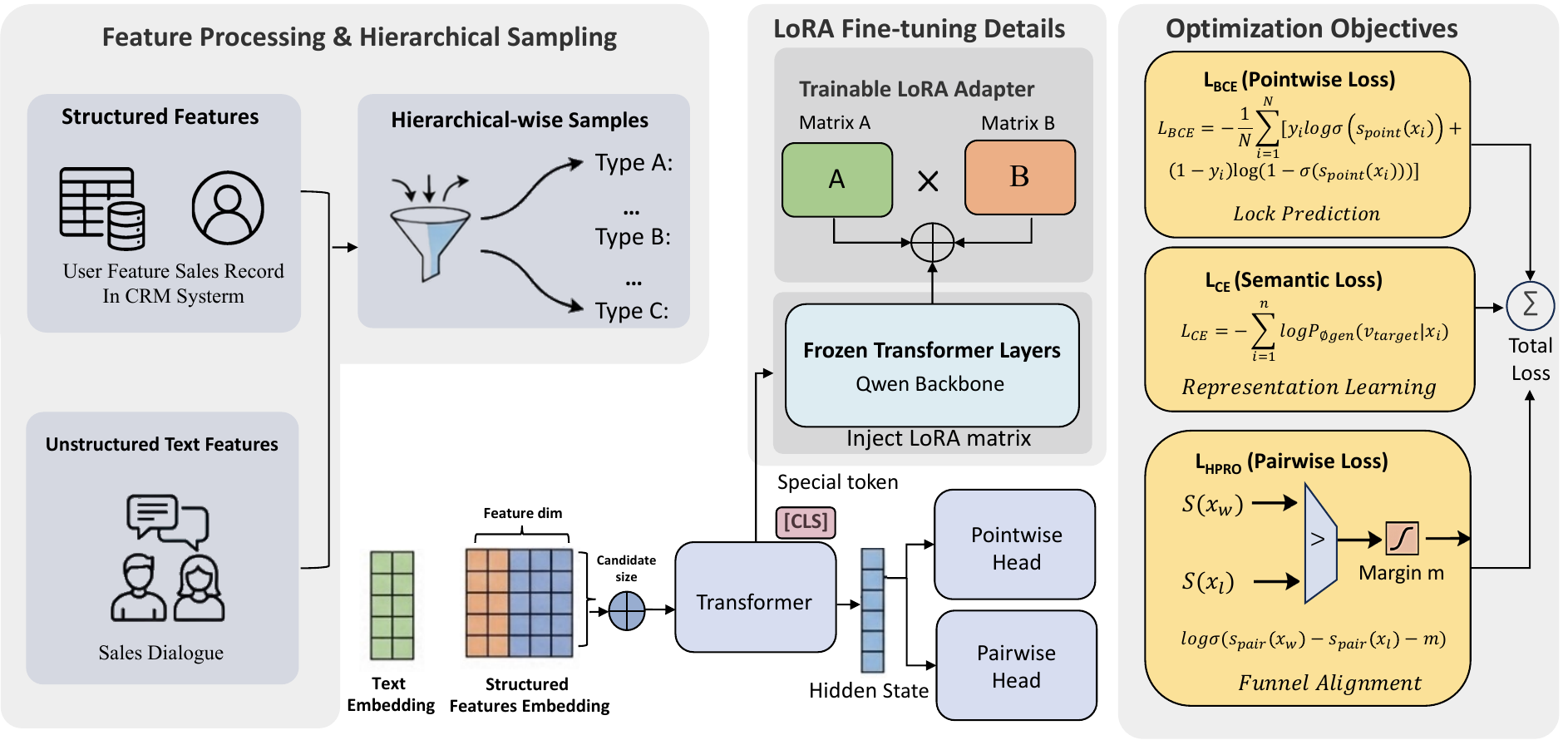}
    \caption{Overview of the proposed framework. Left: Feature processing and hierarchical sampling from structured and unstructured inputs. Middle: LLM backbone with LoRA adaptation. Right: Three optimization objectives corresponding to the triple-head mechanism.}
    \label{fig:architecture}
\end{figure*}


\subsection{Model Architecture}
To bridge semantic understanding with actionable ranking, we propose a discriminative LLM architecture that augments a pre-trained language model with task-specific heads. We term this architecture asLLR (LLM-based Lead Ranking in Auto Sales), illustrated in Figure~\ref{fig:architecture}.



\paragraph{Backbone with LoRA.} We employ a Large Language Model as the encoder to extract high-dimensional semantic representations $\mathbf{h}_{text} \in \mathbb{R}^d$ from the final hidden state. We utilize Low-Rank Adaptation (LoRA)~\cite{hu2022lora} for parameter-efficient fine-tuning.

\paragraph{Triple-Head Mechanism.} Three task-specific heads branch from the backbone's final hidden state:

\begin{enumerate}
    \item \textbf{Semantic Head} ($\phi_{gen}$): The LLM's original vocabulary projection layer ($\mathbb{R}^d \to \mathbb{R}^{|V|}$), retained for next-token prediction. This head serves as a representation regularizer, preventing catastrophic forgetting during domain adaptation.
    
    \item \textbf{Pointwise Head} ($\phi_{point}$): A linear projection ($\mathbb{R}^d \to \mathbb{R}$) outputting a scalar score $s_{point}(x)$ for conversion probability estimation.
    
    \item \textbf{Pairwise Head} ($\phi_{pair}$): A separate linear projection ($\mathbb{R}^d \to \mathbb{R}$) outputting a ranking score $s_{pair}(x)$, optimized via HPRO for relative ordering across funnel stages.
\end{enumerate}

\subsection{Hierarchical Preference Ranking Optimization}
\label{sec:hpro}

Standard pointwise objectives treat all non-converted samples as identical negatives, discarding valuable process information. We propose HPRO to construct dense supervision from funnel hierarchy.

\subsubsection{Funnel-Aware Preference Construction}

We define a funnel hierarchy $\mathcal{F} = \{L_{lock}, L_{drive}, L_{call}, L_{defeat}\}$ ordered by engagement depth, where $L_{lock}$ denotes order lock-in and $L_{defeat}$ denotes lost leads. A mapping $\Phi: \mathcal{X} \to \mathcal{F}$ assigns each sample to its funnel stage.

We construct preference pairs $(x_w, x_l)$ where $\Phi(x_w) \succ \Phi(x_l)$, categorized into three tiers with associated margins:

\begin{itemize}
    \item \textbf{Global Dominance} ($\mathcal{P}_{global}$, margin $m_g$): Lock-in vs. Defeat—the strongest preference signal.
    \item \textbf{Key Action} ($\mathcal{P}_{key}$, margin $m_k$): Test Drive vs. No Drive—a critical intermediate indicator.
    \item \textbf{Soft Signal} ($\mathcal{P}_{soft}$, margin $m_s$): Long Call vs. Short Call—a dense but noisy behavioral signal.
\end{itemize}

\subsubsection{Margin-Aware Bradley-Terry Objective}

HPRO extends the Bradley-Terry model~\cite{bradley1952rank} by injecting funnel-derived margins as probabilistic constraints. For a preference pair $(x_w, x_l)$ with margin $m$, the preference probability is:
\begin{equation}
    P(x_w \succ x_l | m) = \sigma\left(s_{pair}(x_w) - s_{pair}(x_l) - m\right),
\end{equation}
where $\sigma(\cdot)$ is the sigmoid function. The HPRO loss maximizes the log-likelihood:
\begin{equation}
    \mathcal{L}_{HPRO} = -\mathbb{E}_{(x_w, x_l, m) \sim \mathcal{D}_{pair}} \left[ \log \sigma\left(s_{pair}(x_w) - s_{pair}(x_l) - m\right) \right].
\end{equation}

The margin $m$ encodes business priors: larger margins for global dominance pairs enforce clearer separation in the score space, while smaller margins for soft signals allow finer discrimination. In our experiments, we set $m_g = 1.0$, $m_k = 0.5$, and $m_s = 0.1$.

\subsection{Training Objective}

The total objective jointly optimizes all three heads:
\begin{equation}
    \mathcal{L}_{total} = \alpha \mathcal{L}_{BCE} + \lambda_1 \mathcal{L}_{HPRO} + \lambda_2 \mathcal{L}_{CE}.
\end{equation}

\paragraph{Pointwise Loss ($\mathcal{L}_{BCE}$).} Binary cross-entropy for conversion probability calibration:
\begin{equation}
    \mathcal{L}_{BCE} = -\frac{1}{N} \sum_{i=1}^{N} \left[ y_i \log \sigma(s_{point}(x_i)) + (1-y_i) \log (1-\sigma(s_{point}(x_i))) \right].
\end{equation}

\paragraph{Pairwise Loss ($\mathcal{L}_{HPRO}$).} As defined in Eq.~(2), optimizing relative ordering based on hierarchical priors.

\paragraph{Semantic Loss ($\mathcal{L}_{CE}$).} Cross-entropy for QA-style target prediction, preserving the backbone's linguistic capabilities:
\begin{equation}
    \mathcal{L}_{CE} = -\frac{1}{N} \sum_{i=1}^{N} \log P_{\phi_{gen}}(v_{target} | x_i),
\end{equation}
where $v_{target}$ is the target answer token (e.g., ``Yes'' or ``No'' for conversion prediction).


\paragraph{Optimization Strategy.} We observe that purely discriminative losses lead to overfitting; $\mathcal{L}_{CE}$ acts as a regularizer preserving linguistic capability. We apply differential learning rates: task heads use the base rate $5 \times 10^{-5}$, while backbone (LoRA) parameters employ a 20$\times$ lower rate to prevent overfitting. The loss weights $\alpha = 2.0$, $\lambda_1 = 1.0$, and $\lambda_2 = 0.5$ emphasize pointwise calibration while maintaining ranking and regularization objectives.

\section{Experiments}

To validate the proposed HPRO framework, we conducted a multi-stage evaluation progressing from offline benchmarking to mechanism analysis, and finally to real-world deployment.

\subsection{Experimental Setup}
\paragraph{Datasets.}
We evaluate on two proprietary datasets from a leading New Energy Vehicle (NEV) retail system, collected from different operational periods:

\begin{itemize}
    \item \textbf{Benchmark Dataset} (340k samples, 1.45\% positive): An earlier collection enabling fair comparison with CTR baselines under consistent feature engineering.
    \item \textbf{Industrial Dataset} (6.14M samples, 1.33\% positive): A recent full-scale production dataset with richer intermediate behavioral signals (e.g., test drives, call durations) required for training the HPRO module.
\end{itemize}

\noindent Both datasets are split 7:3 (train/test) with strict temporal ordering to prevent look-ahead bias.





\paragraph{Baselines.}
We compare asLLR against six industrial CTR models: Wide\&Deep~\cite{cheng2016wide}, DeepFM~\cite{guo2017deepfm}, xDeepFM~\cite{lian2018xdeepfm}, DCN~\cite{wang2017deep}, DCN-M~\cite{wang2021dcn}, and AutoInt~\cite{song2019autoint}. All baselines share consistent hyperparameters (batch size 256, embedding dimension 8).

\paragraph{Implementation.}
We fine-tune Qwen1.5-1.8B (Scenario 1) and Qwen2.5-1.5B (Scenario 2) with LoRA. Dialogues are truncated to 2,000 tokens. Full hyperparameters and training scripts are provided in our code repository.

\begin{table}[htbp]
  \caption{Results on Benchmark Dataset. Upper: comparison with CTR baselines. $+\mathbf{h}_{text}$ denotes injection of asLLR-extracted textual embeddings. Lower: ablation on asLLR components.}
  \centering
  \begin{tabular}{lccc}
    \toprule
    Model & Embedding & AUC & Gain \\
    \midrule
    \multicolumn{4}{c}{Upper: Comparison with CTR baselines} \\
    \midrule
    W\&D~\cite{cheng2016wide}      & -                    & 0.7860 & - \\
    DeepFM~\cite{guo2017deepfm}    & -                    & 0.7917 & - \\
    xDeepFM~\cite{lian2018xdeepfm} & -                    & 0.7808 & - \\
    DCN~\cite{wang2017deep}        & -                    & 0.7862 & - \\
    DCN-M~\cite{wang2021dcn}       & -                    & 0.7900 & - \\
    AutoInt~\cite{song2019autoint} & -                    & 0.7896 & - \\
    \midrule
    W\&D~\cite{cheng2016wide}      & $+\mathbf{h}_{text}$ & 0.7951 & +0.0091 \\
    DeepFM~\cite{guo2017deepfm}    & $+\mathbf{h}_{text}$ & 0.7976 & +0.0059 \\
    xDeepFM~\cite{lian2018xdeepfm} & $+\mathbf{h}_{text}$ & 0.7895 & +0.0087 \\
    DCN~\cite{wang2017deep}        & $+\mathbf{h}_{text}$ & 0.7911 & +0.0049 \\
    \midrule
    \multicolumn{4}{c}{Lower: Ablation on asLLR components} \\
    \midrule
    asLLR (Base)                 & - & 0.7921 & - \\
    asLLR + $\mathcal{L}_{CE}$   & - & 0.8081 & - \\
    \textbf{asLLR + HPRO (Full)} & - & \textbf{0.8161} & - \\
    \bottomrule
  \end{tabular}
  \label{tab:main_result}
\end{table}
\subsection{Scenario 1: Comparative Benchmarking }
This Scenario investigates whether the LLM backbone can bridge the semantic gap in unstructured CRM logs better than feature interaction-based models. Table~\ref{tab:main_result} presents the results.

\paragraph{Superiority of LLM Backbone.}
The base asLLR model achieves an AUC of 0.7921, surpassing the strongest baseline (DeepFM: 0.7917). This corroborates that pre-trained knowledge within the LLM captures latent customer intents more effectively than shallow embedding lookups.

\paragraph{Embedding Transferability.}
To quantify representation quality, we inject the textual embedding $\mathbf{h}_{text}$ extracted from asLLR into traditional baselines. As shown in Table~\ref{tab:main_result}, this injection yields consistent performance lifts across all baselines (avg. +0.007 AUC), confirming that asLLR learns representations complementary to traditional tabular features.

\paragraph{Ablation on asLLR Variants.}
The lower section of Table~\ref{tab:main_result} isolates component contributions. Adding semantic regularization ($\mathcal{L}_{CE}$) improves AUC to 0.8081, validating that the auxiliary language modeling Scenario prevents catastrophic forgetting. The full model with HPRO achieves \textbf{0.8161}, a +3.0\% improvement over the base model.


\begin{table}[t]
\centering
\caption{Ranking effectiveness on the industrial dataset. P@K\% and R@5.0\% denote precision and recall at top K\% (or 5.0\%) of ranked leads. Relative Lift is computed with respect to asLLR (w/o HPRO).}
\label{tab:task2_industrial}
\begin{tabular}{l|c|cc|c}
\toprule
\multirow{2}{*}{\textbf{Method}} & \multirow{2}{*}{\textbf{AUC}} & \multicolumn{2}{c|}{\textbf{Precision}} & \textbf{Recall} \\
& & \textbf{P@0.1\%} & \textbf{P@1.0\%} & \textbf{R@5.0\%} \\
\midrule
Funnel+Recency & 0.6332 & 3.28\% & 1.50\% &9.23\%  \\
Funnel+CTR  (DeepFM, Two-stage) & 0.6898 & 7.21\% & 1.90\% & 18.76\% \\
Funnel+CTR (DeepFM, Direct)& 0.7382 & 14.41\% & 10.14\% & 21.85\%
\\
\midrule
asLLR (w/o HPRO) & 0.7491 & 18.44\% & 11.56\% & 23.94\% \\
\textbf{asLLR (w/ HPRO)} & \textbf{0.7583} & \textbf{25.76\%} & \textbf{13.33\%} & \textbf{25.18\%} \\
\textit{Relative Lift} & \textit{+1.2\%} & \textit{\textbf{+39.7\%}} & \textit{+15.3\%} & \textit{+5.2\%} \\
\bottomrule
\end{tabular}
\end{table}
\subsection{Scenario 2: Ranking Effectiveness on Industrial Data}
We include three industrial baselines to evaluate different ways of incorporating funnel information into lead prioritization: ``Funnel+Recency'', a rule-based heuristic that ranks leads by funnel stage and then recency; ``Funnel+CTR (DeepFM, Two-stage)'', which first groups leads by funnel stage and then applies DeepFM within each stage; and ``Funnel+CTR (DeepFM, Direct)'', which treats funnel stage as an explicit input feature and directly predicts a ranking score with DeepFM. As shown in Table~\ref{tab:task2_industrial}, simple heuristic ranking is clearly insufficient for this task: Funnel+Recency achieves only 0.6332 AUC and 3.28\% P@0.1\%. Moreover, the two-stage design underperforms direct Funnel+CTR ranking (0.6898 vs. 0.7382 AUC), suggesting that funnel stage should not be imposed as a hard ranking constraint.

Compared with asLLR (w/o HPRO), the full model improves AUC from 0.7491 to 0.7583 and increases P@0.1\% from 18.44\% to 25.76\%, corresponding to a relative improvement of 39.7\%.  It also outperforms Funnel+CTR (DeepFM, Direct), the strongest industrial baseline in our comparison, whose P@0.1\% is 14.41\%.  These gains are especially important in practice because sales teams can only follow up a very limited fraction of leads, making precision at the very top of the ranked list a key determinant of operational efficiency.  Overall, the results suggest that funnel information is most effective when used as structured preference supervision rather than as a rigid ranking rule.
\begin{figure}[h]
\centering
\includegraphics[width=0.8\linewidth]{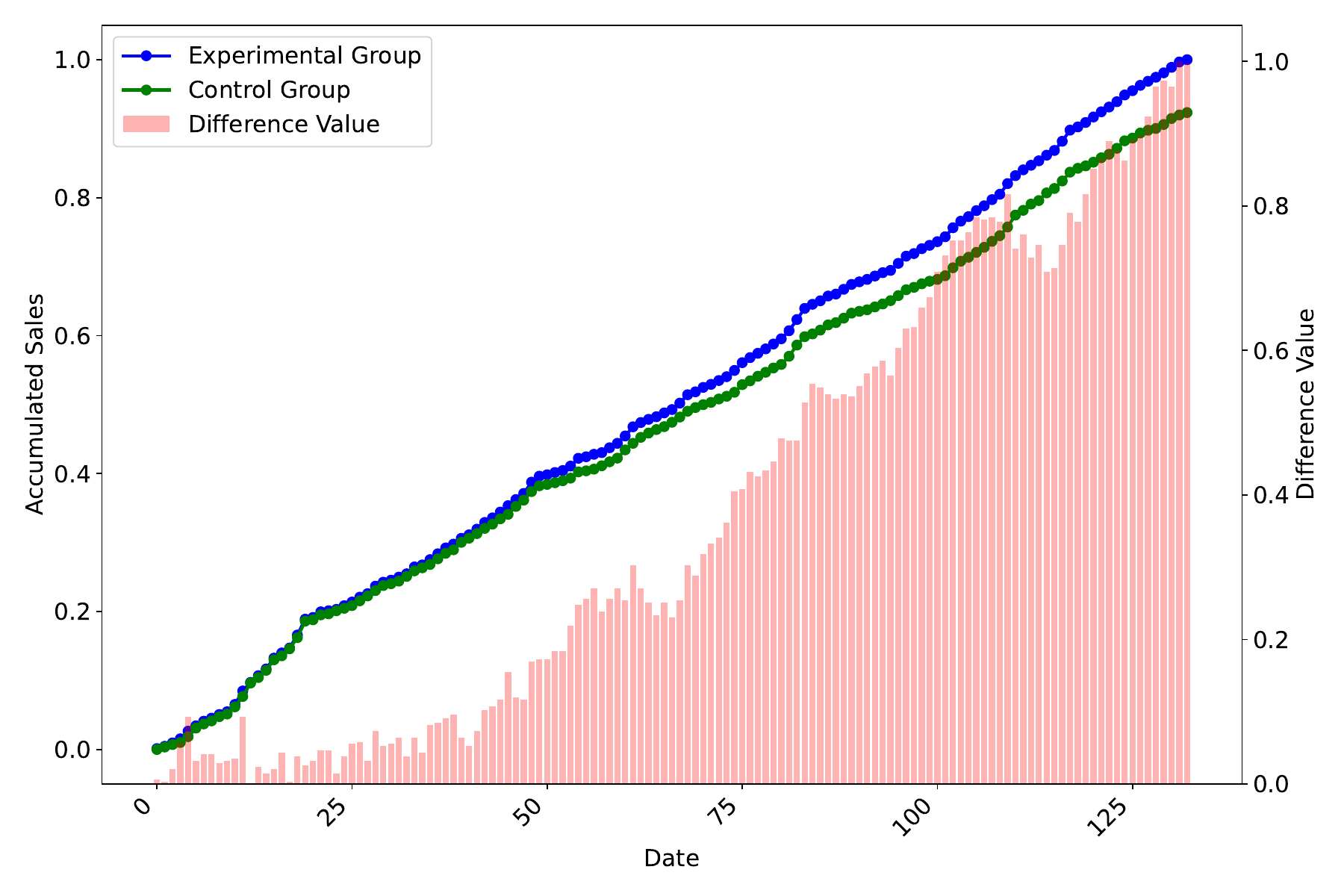}
\vspace{-3mm}
\caption{Cumulative conversion rates (left axis) and relative difference (right axis, shaded) over 132 days.  Stable 9.5\% gap validates HPRO's effectiveness}
\label{fig:result}
\vspace{-3mm}
\end{figure}
\subsection{Online Performance Validation}
\label{sec:task3}

We conducted a province-wide A/B test across all sales operations in a single Chinese province (anonymized) for 132 days. Sales specialists were stratified into two groups with equivalent baselines: a control group using traditional CTR-based lead ranking and an experimental group using HPRO-based recommendations. The experiment employed stratified sampling across four sales capability tiers and 7/14/30-day lock-in counts (lock-ins) as the primary balancing metric and lead follow-up volumes as a secondary factor. Figure \ref{fig:result} shows a 9.5\% relative uplift in lead conversion (two-sided t-test, $p < 0.001$), validating HPRO's real-world sales impact.



\section{Conclusion}
We introduce  a discriminative LLM framework for lead scoring in long-chain sales scenarios. Our key contribution, HPRO, transforms sparse binary supervision into dense, funnel-aware preference signals via a margin-aware Bradley-Terry objective. Experiments on large-scale NEV retail data demonstrate state-of-the-art classification (AUC 0.8161) and ranking performance (+39.7\% P@0.1\%). A 132-day online A/B test validates 9.5\% sales volume uplift, confirming real-world commercial impact.

Although our experiments focus on automotive sales, the proposed framework is applicable to other long-horizon sales domains whenever intermediate customer actions form a meaningful engagement hierarchy.  In such cases, the stage definitions and margin settings can be adapted to domain-specific business processes.

\bibliographystyle{unsrt}  
\bibliography{references}

\end{document}